# High-rectification near-field radiative thermal diode using Weyl semimetals

Yang Hu[1,2,3,#], Haotuo Liu[3,4,#], Bing Yang[5], Kezhang Shi[6], Mauro Antezza[7,8], Xiaohu Wu[3,*], and Yasong Sun[1,2,*]

[1]Basic Research Center, School of Power and Energy, Northwestern Polytechnical University, Xi'an 710072, Shaanxi, P. R. China

[2]Center of Computational Physics and Energy Science, Yangtze River Delta Research Institute of NPU, Northwestern Polytechnical University, Taicang 215400, Jiangsu, P. R. China

[3]Shandong Institute of Advanced Technology, Jinan 250100, Shandong, P. R. China

[4]School of Energy Science and Engineering, Harbin Institute of Technology, Harbin 150001, P. R. China

[5]Centre for Advanced Laser Manufacturing (CALM), School of Mechanical Engineering, Shandong University of Technology, Zibo 255000, P. R. China

[6]Centre for Optical and Electromagnetic Research, National Engineering Research Center for Optical Instruments, Zhejiang University, Hangzhou 310058, P. R. China

[7]Laboratoire Charles Coulomb (L2C) UMR 5221 CNRS-Université de Montpellier, Montpellier F- 34095, France

[8]Institut Universitaire de France, 1 rue Descartes, Paris Cedex 05 F-75231, France

[#]These authors contributed equally: Yang Hu, Haotuo Liu

Email: xiaohu.wu@iat.cn (Xiaohu Wu); yssun@nwpu.edu.cn (Yasong Sun)

**Abstract**

Thermal diodes, which allow heat transfer in a preferential direction while being blocked in a reverse direction, have numerous applications in thermal management, information processing, energy harvesting, etc. Typical materials of thermal diodes in previous works include phase-change and magneto-optical materials. However, such thermal diodes highly depend on specific working temperatures or external magnetic fields. In this work, we propose a near-field radiative thermal diode (NFRTD) based on two Weyl semimetals (WSMs) nanoparticles (NPs) mediated by WSMs planar substrate, which works without external magnetic field and with flexible temperatures. Numerical results show that the maximum rectification ratio of NFRTD can be up to 2673 when the emitter is 200 K and receiver is 180 K, which exceeds the maximum value reported in previous works by more than 10 times. The underlying physical mechanism is the strong coupling of the localized plasmon modes in the NPs and nonreciprocal surface plasmon polaritons in the substrate. In addition, we calculate the distribution of the Green function and reflection coefficient to investigate nonreciprocal energy transfer in NFRTD. Finally, we discuss the effects of momentum-separation on the rectification performance of the NFRTD. This work demonstrates the great potential of WSMs in thermal rectification and paves a novel path in designing high-performance NFRTD.

**1. Introduction**

As one of the most important semiconductor devices, the electric diode is not only of great theoretical significance, but also plays an indispensable role in the modern industry [1]. Inspired by the effective regulation of electric currents by electric diodes, the manipulation of heat flux has also gradually attracted the research interest of related scientists. Recently, the concept of thermal diodes has been proposed [2-9]. The thermal diodes allow heat to transfer in a forward temperature gradient while being blocked in a reverse temperature gradient [10-13]. Owing to this exotic property of directional control of heat flux, it has potential applications in thermal management, thermal rectification, information processing, and energy harvesting [14-19]. To meet the technical requirements of practical applications, thermal diodes often require a rectification ratio of dozens or even higher. So far, thermal diodes based on various energy carriers (photons, phonons, electrons) have been extensively studied [20-26]. However, the rectification ratio of thermal diodes is still inferior to that of typical electrical diodes, which hinders its application and development in related fields.

Due to the coupling effect of evanescent waves, the near-field radiative heat transfer (NFRHT) may exceed the blackbody limit by several magnitude orders [27-46]. The radiative heat flux can be effectively modulated based on the near-field effects, paving a novel path for thermal diodes. Previous works have investigated near-field radiative thermal diode (NFRTD) in a two-body system [2, 47-49], in which the thermal rectification ratio is highly temperature dependent. In 2017, Doyeux et al. first developed the energy transfer in the presence of nonreciprocal object, which suggests that nonreciprocal surface plasmon polaritons (SPPs) can effectively manage the energy transport at the nanoscale [50]. Ott et al. have proposed a NFRTD based on many-body systems [51], in which the NFRHT can be manipulated between nanoparticles (NPs) with nonreciprocity of the magneto-optical materials that depend on the magnetic field. In comparison, the Weyl semimetals (WSMs) can obtain a superior nonreciprocity without an applied magnetic field [52-61]. In addition, the temperature-dependent optical properties of WSMs offer flexibility for the control of the rectification performance of NFRTD. To date, the NFRHT between WSMs has been investigated theoretically [62-64]. However, the rectification potential of the NFRTD based on WSMs has not been explored, and the underlying physical mechanisms need to be further discussed.

In this work, we propose a novel NFRTD using WSMs to achieve a high rectification ratio. The 4×4 transfer matrix method and fluctuational electrodynamics are used to calculate the NFRHT between WSMs NPs mediated by WSMs planar substrate [65, 66]. Owing to strong coupling of the localized plasmon modes in the NPs and nonreciprocal SPPs in the substrate, we can effectively modulate the rectification performance of the NFRTD. In addition, we discuss the distribution of Green function in angular frequency and wavevector space to reveal the underlying physical mechanism. Finally, we investigate the effects of temperature and momentum-separation on the rectification performance of the NFRTD. We believe the results in this work can provide guidance for the design of high-performance NFRTD based on WSMs.

## 2. Modeling and calculation

Schematic diagram of the NFRTD, which consists of a WSMs planar substrate and two WSMs NPs placed above it, is as shown in **Fig. 1**. The radius $R$ of the WSMs NPs is 20 nm. The distance between the NPs and from the NPs to the plate are $d = 1000$ nm and $h = 100$ nm, respectively. The WSMs planar substrate is considered as a semi-infinite medium. The temperature of the red particle (emitter), blue particle (receiver) and planar substrate are $T_1$, $T_2$, and $T_S$, respectively. To simplify the computational model, we assume that $T_1 > T_2 = T_S$. A Cartesian coordinate system is established with the projection of the spherical center of the left particle on the upper surface of the WSMs planar substrate as the origin. When the emitter is on the left side, we define the NFRHT as forward transfer, and the heat flow is $P_1$. The coordinate of emitter and receiver are $\mathbf{r_1} = (0\ 0\ h)$ and $\mathbf{r_2} = (d\ 0\ h)$. When the emitter is on the right side, the NFRHT is backward transfer, and the heat flow is $P_2$. The coordinate of emitter and receiver are $\mathbf{r_1} = (d\ 0\ h)$ and $\mathbf{r_2} = (0\ 0\ h)$. By changing the positions of the WSMs NPs, we can effectively modulate the NFRHT between two WSMs NPs mediated by a WSMs planar substrate and obtain a higher theoretical rectification ratio. It is worth noting that the dipole approximation can be nicely adapted when $d$, $h$, $\lambda \gg R$, two NPs can be described as point sources [66, 67].

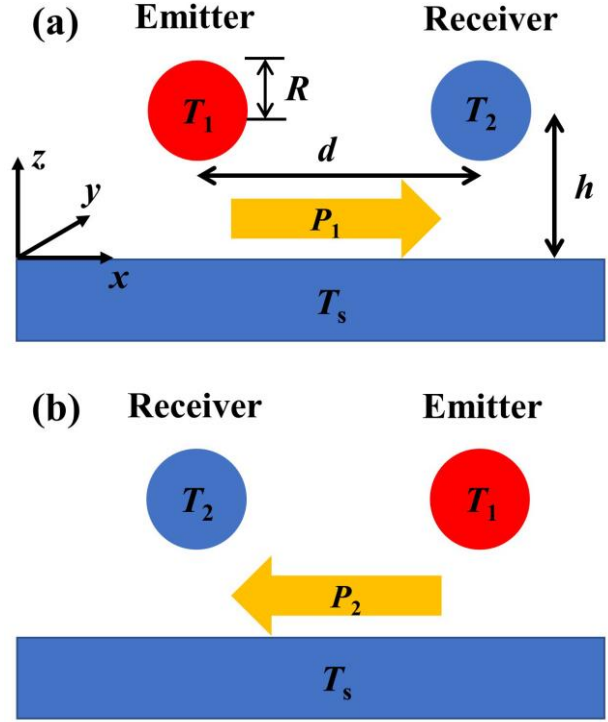

**Fig. 1** Schematic diagram of the NFRTD. The two NPs with radius $R$ are placed along the $x$-axis at a gap distance of $d$ and in the vicinity of a WSMs substrate at a distance of $h$. The coordinate of the emitter and receiver are $\mathbf{r}_1$ and $\mathbf{r}_2$. (a) $\mathbf{r}_1 = (0\ 0\ h)$, $\mathbf{r}_2 = (d\ 0\ h)$, and (b) $\mathbf{r}_1 = (d\ 0\ h)$, $\mathbf{r}_2 = (0\ 0\ h)$ respectively. The temperature of the emitter, receiver, and substrate are $T_1$, $T_2$, and $T_s$. The forward and backward heat flux are $P_1$ and $P_2$, respectively.

Specifically, we considered the simplest case of WSM. The time-reversal symmetry could be broken when splitting a Dirac point into a pair of Weyl nodes with opposite chirality. Each pair of Weyl nodes is separated in momentum space by wavevector **b**. This kind of material has been recently realized experimentally, such as EuCd$_2$As$_2$. Moreover, the presence of Weyl nodes changes the electromagnetic response, and the displacement electric field for WSM can be written as [52, 53]

$$\mathbf{D} = \varepsilon_d \mathbf{E} + \frac{ie^2}{4\pi^2 \hbar \omega}\left(-2b_0 \mathbf{B} + 2\mathbf{b}\times\mathbf{E}\right), \tag{1}$$

where $e$ is the elementary charge, $\hbar$ is the reduced Planck constant, **E** is the electric field, **B** is the magnetic flux density, and $\varepsilon_d$ is the permittivity of the corresponding Dirac semimetal. It is worth noting that Dirac semimetals are generally assumed to be isotropic in the absence of external magnetic fields. Thus, we assume that the diagonal elements of the permittivity tensor are all $\varepsilon_d$. The first (-2$b_0$**B**) and the second (2**b**×**E**) terms in the parentheses of Eq. (1) describe the chiral magnetic effect and the anomalous Hall effect, respectively. In this work, we only consider materials where the Weyl nodes have the same energy (i.e., $b_0 = 0$). The momentum-separation **b** of the Weyl nodes is an axial vector that acts similar to an internal magnetic field, and we choose the coordinates with **b** along the positive $y$-direction: (i.e., **b** = $b$**y**). With the above considerations, the permittivity tensor of the WSM becomes

$$\boldsymbol{\varepsilon} = \begin{bmatrix} \varepsilon_d & 0 & i\varepsilon_a \\ 0 & \varepsilon_d & 0 \\ -i\varepsilon_a & 0 & \varepsilon_d \end{bmatrix}, \qquad (2)$$

where

$$\varepsilon_a = \frac{be^2}{2\pi^2 \varepsilon_0 \hbar \omega}. \qquad (3)$$

When $b \neq 0$, $\varepsilon_a$ is nonzero. Thus, $\varepsilon$ is asymmetric and could break Lorentz reciprocity. To calculate the diagonal term $\varepsilon_d$, we apply the Kubo Greenwood formalism within the random phase approximation to a two-band model with spin degeneracy. This formalism considers both interband and intraband transitions [68, 69]

$$\varepsilon_d = \varepsilon_b + i\frac{\sigma}{\omega}, \qquad (4)$$

where σ is the bulk conductivity given by:

$$\sigma = \frac{r_s g}{6}\Omega G\left(\frac{\Omega}{2}\right) + i\frac{r_s g}{6\pi}\left\{\frac{4}{\Omega}\left[1 + \frac{\pi^2}{3}\left(\frac{k_B T}{E_F(T)}\right)^2\right] + 8\Omega \int_0^{\xi_c} \frac{G(\xi) - G\left(\frac{\Omega}{2}\right)}{\Omega^2 - 4\xi^2}\xi d\xi\right\}, \quad (5)$$

where $\varepsilon_b$ is the background permittivity, $\Omega = \hbar(\omega + i\tau^{-1})/E_F$ is the complex frequency normalized by the chemical potential, $\tau^{-1}$ is the scattering rate corresponding to Drude damping, $G(E) = n(-E) - n(E)$ where $n(E)$ is the Fermi distribution function, $E_F(T)$ is the chemical potential, $r_s = e^2/4\pi\varepsilon_0\hbar v_F$ is the effective fine structure constant, $v_F$ is the Fermi velocity, $g$ is the number of Weyl points, $\xi_c = E_c/E_F$ where $E_c$ is the cutoff energy beyond which the band dispersion is nonlinear. Other important parameters are as follows: $\varepsilon_b$ = 6.2, $\xi_c$ = 3, $\tau$ = 1000 fs, $g$ = 2, and $v_F$ = 0.83×10$^9$ m/s. The chemical potential as a function of temperature can be calculated from charge conservation [52]:

$$E_F(T) = \frac{2^{1/3}\left[9E_F(0)^3 + \sqrt{81E_F(0)^6 + 12\pi^6 k_B^6 T^6}\right]^{2/3} - 2\pi^2 3^{1/3} k_B^2 T^2}{6^{2/3}\left[9E_F(0)^3 + \sqrt{81E_F(0)^6 + 12\pi^6 k_B^6 T^6}\right]^{1/3}}, \quad (6)$$

where $E_F$ (0 K) = 0.163 eV and $E_F$ (300 K) = 0.150 eV.

The permittivity of magneto-optical material InSb is given by [70]:

$$\boldsymbol{\varepsilon} = \begin{bmatrix} \varepsilon_1 & 0 & i\varepsilon_2 \\ 0 & \varepsilon_3 & 0 \\ -i\varepsilon_2 & 0 & \varepsilon_1 \end{bmatrix}. \quad (7)$$

Note that the direction of the magnetic field is in the positive $y$ direction. The detailed calculation parameters can be found in Ref. [71].

The permittivity ratio represents the ratio of the non-diagonal elements to the main diagonal elements in the permittivity tensor, which is an effective parameter to describe the performance of nonreciprocal materials [69]. To compare the nonreciprocity of the two materials, we define the permittivity ratio: $|\varepsilon_a/\varepsilon_d|$ for WSMs and $|\varepsilon_1/\varepsilon_2|$ for InSb. **Fig. 2**(a) shows the permittivity ratio of WSMs and Insb as a function of angular frequency. Note that the temperature of WSMs is 300 K. The off-diagonal elements of the WSMs and InSb are determined by momentum-separation $b$ and magnetic field, respectively. It can be seen that even when InSb is under a very strong magnetic field (~ 2 T), its nonreciprocity is still inferior to that of WSMs. This is due to the separation of the Weyl nodes in momentum-separation $b$ having relatively large values for compounds such as $EuCd_2As_2$. In addition, the strong nonreciprocity of WSMs can be obtained without an applied magnetic field, which is its significant advantage over the magneto-optical material InSb. We further calculated the WSMs permittivity tensor components ($\varepsilon_a$ and $\varepsilon_d$) versus angular frequency, as shown in **Fig. 2**(b). We find that $\varepsilon_a$ is comparable to $\varepsilon_d$ over the entire display wavelength range. In addition, $\varepsilon_a$ and $\varepsilon_d$ are strongly correlated with the momentum-separation and temperature, respectively, which provides us with more possibilities for the nonreciprocal regulation of WSMs.

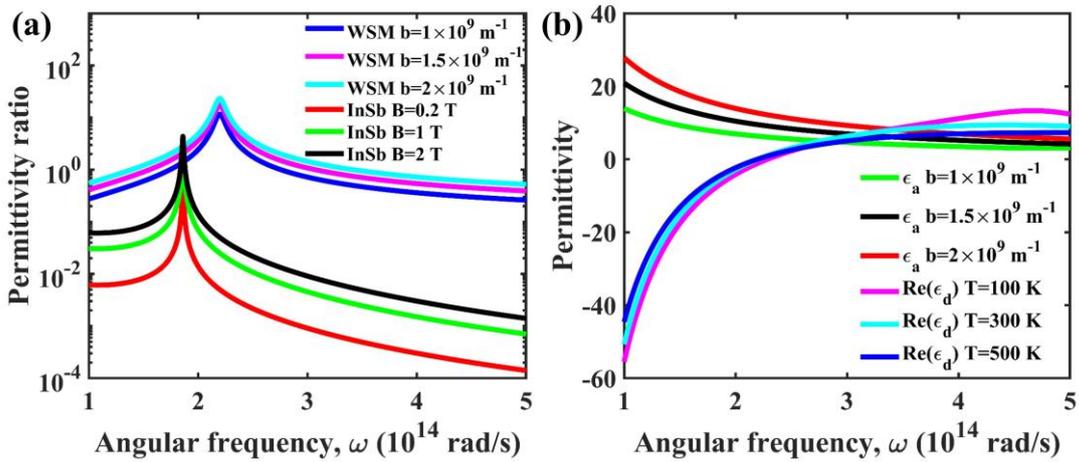

**Fig.2** (a) The permittivity ratio ($|\varepsilon_a/\varepsilon_d|$ for WSMs and $|\varepsilon_1/\varepsilon_2|$ for InSb) varies with angular frequency. The temperature of WSMs is 300 K. (b) The permittivity tensor components ($\varepsilon_a$ and $\varepsilon_d$) of WSMs vary with angular frequency.

The total heat flux received by NP$_i$ can be described as [51]

$$P_i = \int_0^\infty \frac{d\omega}{2\pi} P_{i,\omega} = 3\int_0^\infty \frac{d\omega}{2\pi}[\Theta(T_j) - \Theta(T_i)]\tau_{ji}, \qquad (8)$$

where $\Theta(T) = \hbar\omega/[\exp(\hbar\omega/k_B T) - 1]$ is the mean energy of the Planck thermal harmonic oscillators, and $\tau_{ji} = 4k_0^4 Tr[\chi_i \mathbf{G}_{ij} \chi_j \mathbf{G}_{ij}^*]/3$ is the transmission coefficient for the heat flux between the NPs [72, 73]. Here * denotes conjugate transpose. $\chi_i = (\alpha_i - \alpha_i^*)/2i$ is the dressed polarizability of NP$_i$ [67, 72]. When ignoring radiative correction, the polarizability of an anisotropic NP could be written in Clausius-Mossoti form [32, 74]. For dielectric NP, the magnetic polarizability can be ignored. When NP is surrounded by vacuum, the electric polarizability of the anisotropic NPs $\boldsymbol{\alpha}_i$ is given as [32]

$$\boldsymbol{\alpha}_i = 4\pi R^3 \frac{\boldsymbol{\varepsilon}_i - 1\mathbf{I}}{\boldsymbol{\varepsilon}_i + 2\mathbf{I}}, \qquad (9)$$

where $\mathbf{I}$ is the unit matrix, $\boldsymbol{\varepsilon}_1$ ($\boldsymbol{\varepsilon}_2$) is the permittivity tensor of the emitter (receiver). $\mathbf{G}_{ij}$ in $\tau_{ji}$ is the dyadic Green tensor:

$$\mathbf{G}_{ij} = \mathbf{G}_{ij}^{(0)} + \mathbf{G}_{ij}^{(sc)}, \qquad (10)$$

where $\mathbf{G}_{ij}^{(0)}$ is vacuum Green function, which is related to the position of NPs [75]

$$\mathbf{G}_{ij}^{(0)} = \mathbf{G}^{(0)}(\mathbf{r}_i, \mathbf{r}_j) = \frac{e^{ik_0 r_{ij}}}{4\pi r_{ij}}\left[\left(1 + \frac{ik_0 r_{ij} - 1}{k_0^2 r_{ij}^2}\right)\mathbf{I} + \frac{3 - 3ik_0 r_{ij} - k_0^2 r_{ij}^2}{k_0^2 r_{ij}^2}\hat{\mathbf{r}}_{ij} \otimes \hat{\mathbf{r}}_{ij}\right], \qquad (11)$$

where $k_0$ is the vacuum wavevector, $r_{ij} = |\mathbf{r}_{ij}|$ is the magnitude of the vector linking two NPs, $\hat{\mathbf{r}}_{ij} = \mathbf{r}_{ij}/r_{ij}$ and $\otimes$ is the outer product.

$$\mathbf{G}_{ij}^{(sc)} = \frac{i}{8\pi^2}\int_{-\infty}^\infty dk_x \int_{-\infty}^\infty \left(r_{ss}\mathbf{M}_{ss} + r_{ps}\mathbf{M}_{ps} + r_{sp}\mathbf{M}_{sp} + r_{pp}\mathbf{M}_{pp}\right)e^{i\left[k_x(x_i - x_j) + k_y(y_i - y_j)\right]}e^{ik_z|z_i + z_j|}dk_y \qquad (12)$$

where $\mathbf{G}_{ij}^{(sc)}$ is scatter Green function, which is associated with substrate presence

[76]. *r* is the reflection coefficient related to incident "*s*" and "*p*" polarized waves. $k_x$ and $k_y$ are the in-plane wavevectors, and $k_z$ is the out-plane wavevector.

$$\mathbf{M}_{ss} = \frac{1}{k_z k_\rho^2} \begin{pmatrix} k_y^2 & -k_x k_y & 0 \\ -k_x k_y & k_x^2 & 0 \\ 0 & 0 & 0 \end{pmatrix}, \mathbf{M}_{pp} = \frac{k_z}{k_0^2 k_\rho^2} \begin{pmatrix} -k_x^2 & -k_x k_y & -k_x k_\rho^2/k_z \\ -k_x k_y & -k_y^2 & -k_y k_\rho^2/k_z \\ k_x k_\rho^2/k_z & k_y k_\rho^2/k_z & k_\rho^4/k_z^2 \end{pmatrix},$$

$$\mathbf{M}_{sp} = \frac{1}{k_0 k_\rho^2} \begin{pmatrix} -k_x k_y & -k_y^2 & -k_y k_\rho^2/k_z \\ k_x^2 & k_x k_y & k_x k_\rho^2/k_z \\ 0 & 0 & 0 \end{pmatrix}, \mathbf{M}_{ps} = \frac{1}{k_0 k_\rho^2} \begin{pmatrix} k_x k_y & -k_x^2 & 0 \\ k_y^2 & -k_x k_y & 0 \\ -k_y k_\rho^2/k_z & k_x k_\rho^2/k_z & 0 \end{pmatrix}. \quad (13)$$

## 3. Results and discussion

To quantitatively describe the performance of the NFRTD, we define the rectification ratio as the ratio of the difference between the forward and backward total heat flux to the minimum total heat flux [49, 77]:

$$\eta = \frac{P_2 - P_1}{P_1}, \quad (14)$$

**Fig. 3** gives the relationship of the rectification ratio on the temperature of the emitter ($T_1$) and receiver ($T_2$), where $T_1 > T_2$. The temperature of the planar substrate is equal to that of the receiver, i.e., $T_2 = T_S$. The momentum-separation *b* is set to $2\times10^9$ m$^{-1}$. We found that the rectification ratio strongly depends on the temperature with WSMs NPs. When $T_1$ = 200 K and $T_2$ = 180 K, we can obtain a rectification ratio of 2673, which exceeds the maximum rectification ratio of previous works by more than a factor of 10 [51, 71, 76, 78]. The low-temperature thermal diodes have applications in deep and low-temperature heat harvesting and heat transfer technology [79-81]. This may be due to the effect of temperature on the nonreciprocity of the WSMs, resulting in a significant difference between the forward and backward heat flux. In addition, it could achieve a high rectification ratio at a smaller temperature difference compared to conventional

NFRTD.

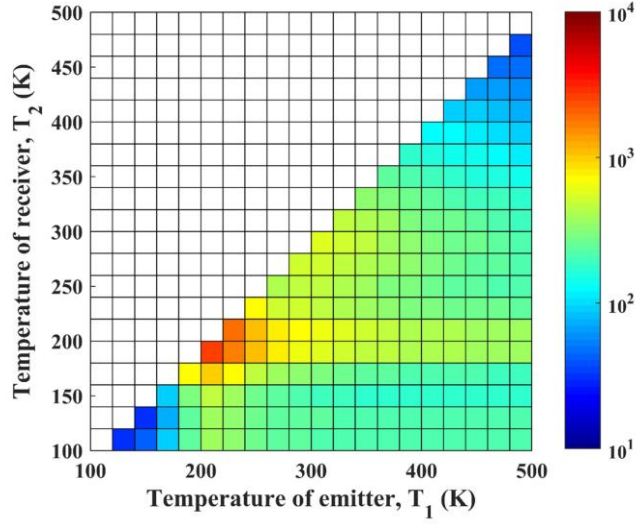

**Fig. 3** The rectification ratio varies with the temperature of the emitter $T_1$ and the temperature of the receiver $T_2$. The temperature of the planar substrate is $T_S = T_2$. The momentum-separation is $b = 2\times10^9$ m$^{-1}$.

To understand the high rectification ratio, we calculated the forward transmission coefficient $T_{12}$ and backward transmission coefficient $T_{21}$ between WSMs NPs. As shown in **Fig. 4**(a), three resonances occur at $1.35\times10^{14}$ rad/s, $2.1\times10^{14}$ rad/s, and $3.1\times10^{14}$ rad/s for both $T_{12}$ and $T_{21}$. The enhancement of NFRHT is related to the coupling of threefold degenerate localized modes in NPs and the surface modes in the substrate. The frequencies of localized plasmon modes $\omega_{m=0,\pm1}$ corresponding to quantum number m = 0, ±1 is determined by dressed polarizability, which can be expressed as [82]:

$$\chi_i = \begin{pmatrix} \chi_{11} & 0 & \chi_{13} \\ 0 & \chi_{22} & 0 \\ \chi_{31} & 0 & \chi_{33} \end{pmatrix}, \tag{15}$$

where $\chi_{11} = \chi_{33}$, and $\chi_{13} = -\chi_{31}$. The resonance frequency of WSMs NPs can be

determined by [83-85]:

$$\det(\boldsymbol{\varepsilon}(\omega,T)+2\mathbf{I})=(\varepsilon_d+2)((\varepsilon_d+2)^2-\varepsilon_a^2)=0, \tag{16}$$

Thus, the resonance modes appear when $(\varepsilon_d+2)=0$ or $(\varepsilon_d+2)^2-\varepsilon_a^2=0$. For the NFRHT between WSMs NPs, the main diagonal component of the dressed polarizability tensor dominates. Next, we mainly analyzed these three components. The resonance of $1.35\times10^{14}$ rad/s, $2.1\times10^{14}$ rad/s, and $3.1\times10^{14}$ rad/s correspond to $\chi_{11}$, $\chi_{22}$, and $\chi_{33}$, respectively. The resonance of $2.1\times10^{14}$ rad/s depends on the localized plasmon mode of $(\varepsilon_d+2)=0$, thus it is strongly related to the temperature of WSMs NPs. This momentum-separation independent mode is the "non-rotating" mode that exhibits zero angular momentum (m = 0). The $T_{12}$ is several orders of magnitude lower than $T_{21}$, demonstrating strong nonreciprocity due to the temperature difference of WSMs NPs. The resonance of $1.35\times10^{14}$ rad/s, and $3.1\times10^{14}$ rad/s depend on the localized circular mode of $(\varepsilon_d+2)^2-\varepsilon_a^2=0$, thus it is related to both temperature and momentum-separation of WSMs NPs. The resonance at angular frequency of $3.1\times10^{14}$ rad/s for the heat flux rotates clockwise along the momentum-separation (m = -1). The $T_{12}$ and $T_{21}$ are equal, and there is no nonreciprocity. The resonance at angular frequency of $1.35\times10^{14}$ rad/s for the heat flux rotates counter-clockwise along the momentum-separation (m = +1). The $T_{12}$ is a little larger than $T_{21}$, exhibiting weaker nonreciprocity. **Figs. 4**(b) and (d) give the spectral heat flux and dressed polarizability of WSMs NPs for the momentum-separation of $b = 0$. Only the localized plasmon modes are excited, and the NFRHT between the WSMs NPs becomes strictly reciprocal in this case. Moreover, the polarizability tensor has only equal main diagonal elements.

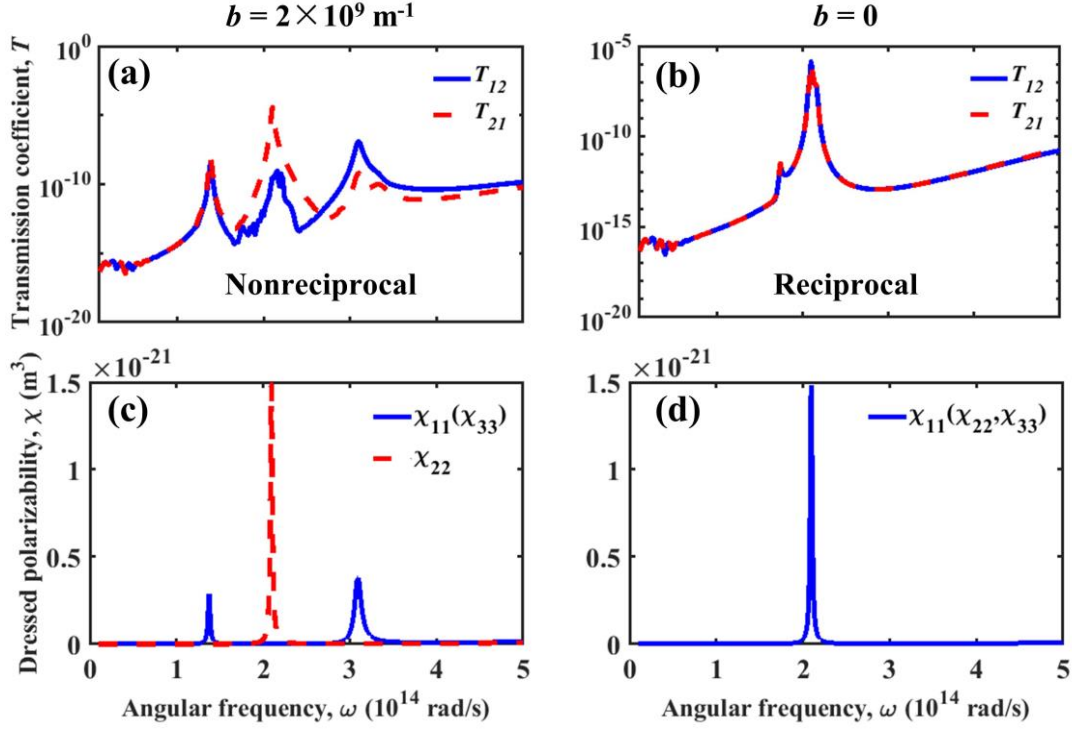

**Fig. 4** The forward transmission coefficient $T_{12}$ and backward transmission coefficient $T_{21}$ between WSMs NPs for the momentum-separation of (a) $b = 2\times10^9$ m$^{-1}$ and (b) $b = 0$. The temperature of the emitter and receiver are $T_1 = 200$ K, and $T_2 = 180$ K. The dressed polarizability of WSMs NPs at temperature of 200 K for the momentum-separation of (c) $b = 2\times10^9$ m$^{-1}$ and (d) $b = 0$.

To explain the heat transfer mechanism, the Green function for the forward and backward heat flux varying with angular frequency are investigated (**Fig. 5**). Since the nonreciprocal effect of this system depends mainly on the substrate, the polarizability effect on the Green function is not considered. Eq. (10) indicates that Green function **G** consists of a vacuum function $\mathbf{G}^{(0)}$ and a scattering function $\mathbf{G}^{(sc)}$. The Green function in the transmission coefficient can be decomposed into three parts:

$$\mathrm{Tr}(\mathbf{GG}^*) = \mathrm{Tr}(\mathbf{G}^{(0)}\mathbf{G}^{(0)*}) + \mathrm{Tr}(\mathbf{G}^{(sc)}\mathbf{G}^{(sc)*}) + 2\mathrm{Tr}(\chi\mathbf{G}^{(0)}\chi\mathbf{G}^{(sc)*}), \qquad (17)$$

the right side of Eq. (17) represents the sum of vacuum, scattering, and cross contributions [86], respectively. The scattering and cross contribution are associated with the presence of the substrate. The $\mathbf{G}^{(sc,sc)}$ agrees well with the $\mathbf{G}$, representing that the NFRHT is mainly through the substrate rather than vacuum. When $b = 0$, the Green function varies with different angular frequencies (**Fig. 5 (a)**). The total Green function (black line) has only a resonance at the angular frequency of $2.17 \times 10^{14}$ rad/s. At this angular frequency, the contribution of the total Green function is mainly by the substrate, and the frequency corresponds to the position where Re($\varepsilon_d$) is -1, indicating the excitation of surface modes in the substrate enhances the NFRHT. While when $b = 2 \times 10^9$ m$^{-1}$, the NFRHT has nonreciprocity, and the green function $G_{1,\omega}$ and $G_{2,\omega}$ corresponding to $P_1$ and $P_2$ are shown in **Fig. 5 (b)** and **Fig. 5 (c)**, respectively. For $b = 2 \times 10^9$ m$^{-1}$, three different resonances occur at $2.17 \times 10^{14}$ rad/s, $1.4 \times 10^{14}$ rad/s, and $3.02 \times 10^{14}$ rad/s due to the non-zero off-diagonal permittivity elements. The resonance at $1.4 \times 10^{14}$ rad/s is small and has a negligible effect on the total Green function. The resonance at $2.17 \times 10^{14}$ rad/s in $G_{1,\omega}$ is much smaller than that in $G_{2,\omega}$, as affected by the substrate. Therefore, the directional nonreciprocal surface modes in substrate leading to the spectral heat flux in $P_{1,\omega}$ is smaller than $P_{2,\omega}$. At the frequency of $3.02 \times 10^{14}$ rad/s, the $G_{1,\omega}$ is larger than $G_{2,\omega}$ corresponding to $P_{1,\omega}$ is larger than $P_{2,\omega}$. The mechanism is also due to the excitation of nonreciprocal surface modes, while the nonreciprocal effect is weaker.

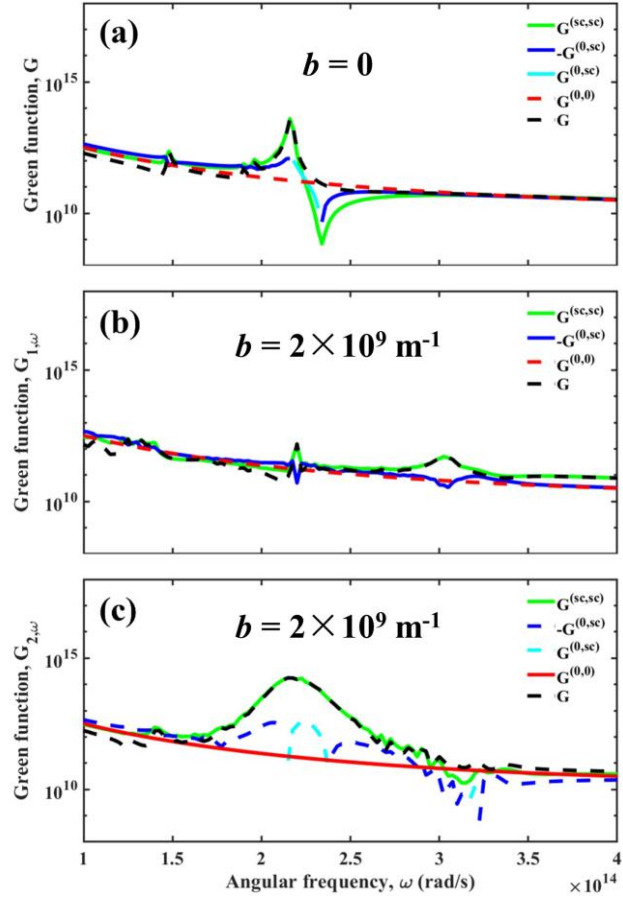

**Fig. 5** The Green function varies with angular frequency. Green lines represent the scattering contribution, blue lines represent the negative cross contribution, cyan lines represent the positive cross contribution, red lines represent the vacuum contribution, and black lines represent the total contribution. (a) $b = 0$, (b) $b = 2\times10^9$ m$^{-1}$ corresponding to $P_1$, and (c) $b = 2\times10^9$ m$^{-1}$ corresponding to $P_2$.

Further, we calculate the distribution of the scattering Green function in the wavevector space to obtain an intuitive understanding of the influence of the substrate on the nonreciprocal energy transfer, as shown in **Fig. 6**. When $b = 0$, the scattering Green function is consistent on the forward and backward heat flux, thus there is no nonreciprocal effect in the near-field thermal diode (**Figs. 6** (a), (b)). When the momentum-separation $b = 2\times10^9$ m$^{-1}$, the symmetry of eigenstates with positive and negative $k_x$ is damaged by the momentum-separation, leading to a strong nonreciprocal surface mode in the substrate. When the angular frequency $\omega = 1.35\times10^{14}$ rad/s, the forward and reverse heat fluxes are almost equal. Therefore, it can be considered that the NFRTDs are reciprocal, which reflects a good agreement with **Fig. 4** (a). When the angular frequency $\omega = 2.1\times10^{14}$ rad/s, the distribution of scattering Green function in positive and negative $k_x$ space has an asymmetrical character. Specifically, the narrow blue bright band at the positive $k_x$ space shifts to red from **Fig. 6** (e) to **Fig. 6** (f), which significantly enhances the radiative heat flux $P_2$, producing a strong nonreciprocity. When the angular frequency $\omega = 3.1\times10^{14}$ rad/s, the distribution of scattering Green function in negative $k_x$ space is slightly different in **Fig. 6** (g) and **Fig. 6** (h), leading to smaller nonreciprocity.

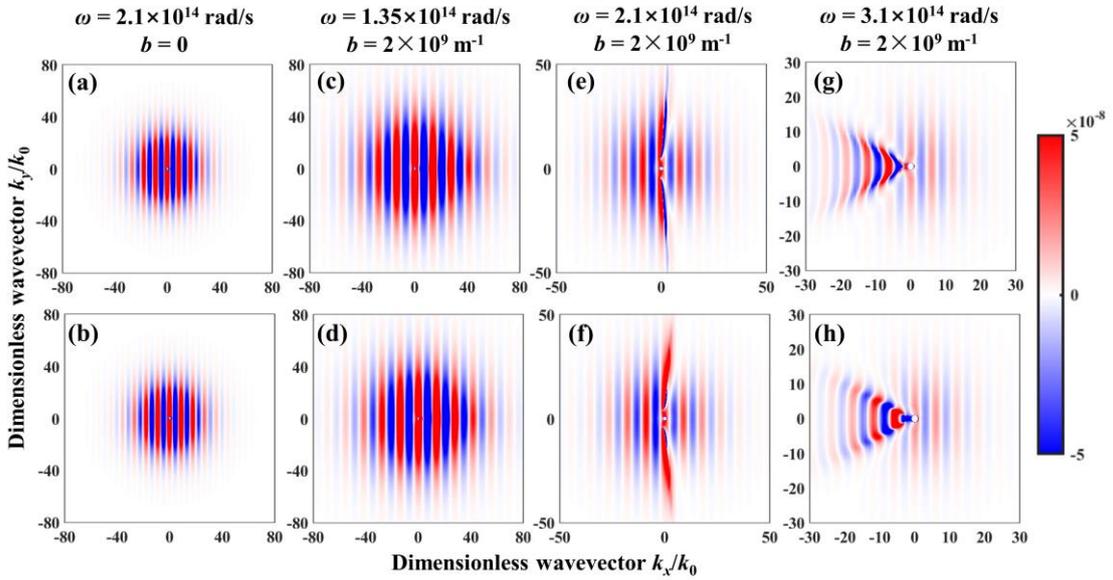

**Fig. 6** The real part of $G^{(sc)}(3,3)$ varies with dimensionless wavevector components $k_x$ and $k_y$ for different for angular frequencies and momentum-separations: (a), (b) $\omega = 2.1\times10^{14}$, $b = 0$, (c), (d) $\omega = 1.35\times10^{14}$, $b = 2\times10^9$ m$^{-1}$, (e), (f) $\omega = 2.1\times10^{14}$, $b = 2\times10^9$ m$^{-1}$, and (g), (h) $\omega = 3.1\times10^{14}$, $b = 2\times10^9$ m$^{-1}$. The first row corresponds to $P_{1,\omega}$, and the second row corresponds to $P_{2,\omega}$.

To explain the mechanism of the asymmetric distribution of Green function at different resonance frequencies, the imaginary part of $r_{pp}$ is shown in **Fig. 7**. The two "$p$" in the subscript of $r_{pp}$ indicate the incoming $p$-polarized wave and reflected $p$-polarized wave, respectively. The imaginary part of $r_{pp}$ is proportional to the density of states of the photons, and the near-field coupling of the two WSMs NPs via the nonreciprocal SPPs of the substrate is mainly given by the component $r_{pp}$ of the reflection tensor $\begin{pmatrix} r_{pp} & r_{ps} \\ r_{sp} & r_{ss} \end{pmatrix}$ [71, 87]. Thus, we can understand the underlying physical mechanisms by focusing on the component $r_{pp}$. When the momentum-separation **b** = 0, the WSMs substrate supports isotropic surface modes because there is no off-diagonal component in the permittivity tensor. While when the momentum-separation $b = 2 \times 10^9$ m$^{-1}$, the momentum-separation is along the positive $y$ direction, the reflection coefficient shows asymmetry in positive and negative $k_x$ spaces. At angular frequency of $1.35 \times 10^{14}$ rad/s, the value of the reflection coefficient is small, and the non-reciprocal effect is weak (Figure. 7b). However, **Fig. 7** (c) shows two narrow bright bands in positive $k_x$ space at angular frequency of $2.1 \times 10^{14}$ rad/s, leading to a strong non-reciprocal effect. When the angular frequency $\omega = 3.1 \times 10^{14}$ rad/s, the reflection coefficient has a wide bright band but a small value, resulting in a weak non-reciprocal effect.

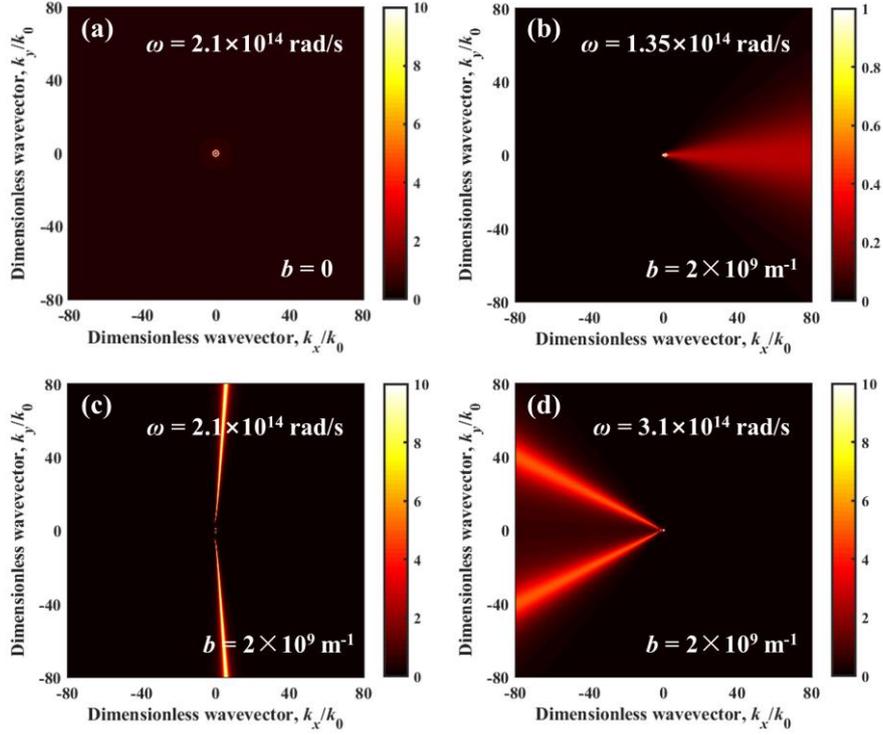

**Fig. 7** The imaginary part of $r_{pp}$ varies with wavevector components $k_x$ and $k_y$ for different for angular frequencies and momentum-separations: (a) $\omega=1.35\times10^{14}$ rad/s, $b=0$, (b) $\omega=1.35\times10^{14}$ rad/s, $b=2\times10^9$ m$^{-1}$ (c) $\omega=2.1\times10^{14}$ rad/s, $b=2\times10^9$ m$^{-1}$, and (d) $\omega=3.1\times10^{14}$ rad/s, $b=2\times10^9$ m$^{-1}$.

Our study is not limited to a specific material, but a class of materials with different momentum-separations. We further investigate the rectification ratio $\eta$ versus momentum-separation **b** and give the corresponding permittivity ratio $|\varepsilon_a/\varepsilon_d|$ to discuss the performance of the NFRTD, as shown in **Fig. 8**. The temperature of emitter and receiver as $T_1 = 200$ K and $T_2 = 180$ K, respectively. The rectification ratio $\eta$ fluctuates upward with the increase of momentum separation. The permittivity ratio $|\varepsilon_a/\varepsilon_d|$ increases with increasing momentum-separation at the resonance frequency of localized plasmon modes $\omega=2.1\times10^{14}$ rad/s when the temperature is 200 K. Therefore, the nonreciprocal effect of the NFRTD was enhanced, leading to a higher rectification ratio. When the momentum-separation $b = 2\times10^9$ m$^{-1}$, the rectification ratio can reach 2673. So far, the rectification ratio is ∼ 249 for InSb NPs mediated by InSb substrate, and ∼ 88.67 for SiC NPs mediated by drift-biased graphene grating [71, 78]. The strong rectification of conventional thermal diodes, such as intrinsic semiconductors or vanadium dioxide-based rectification method, occurs only close to the critical temperature of these materials. Furthermore, obtaining a good rectification effect often requires a large temperature difference, which is difficult to implement experimentally. In contrast, WSM-based near-field radiative diode could get a high rectification ratio in a small temperature difference. In addition, the rectification effect varies with the emitter and receiver could change with the parameters of WSM, such as the number of Weyl points $g$, the Fermi velocity $v_F$, the background permittivity $\varepsilon_b$, and the scattering rate corresponding to Drude damping $\tau$.

Our findings can provide theoretical guidance for multi-probe based near-field thermal radiation systems similar to the model of Ben-Abdallah [88]. A direct measurement of the rectification effect might be possible with a many-body heat transfer setup like that developed by Reddy's group seems to be feasible [89]. There may be other possibilities, but since our work is biased toward theoretical analysis, it is necessary to explore more details and make further speculations before we move toward experiments.

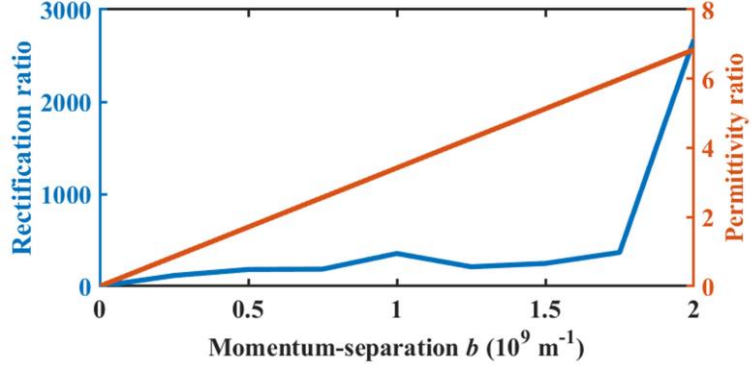

**Fig. 8** Left: the rectification ratio $\eta$ varying with momentum-separation. The temperature of the emitter is 200 K, and the temperature of the receiver is 180 K. Right: permittivity ratio $|\varepsilon_a/\varepsilon_d|$ varying with momentum-separation when the temperature is 200 K, and angular frequency is $2.1\times10^{14}$ rad/s.

## 4. Conclusion

In summary, we theoretically investigate a novel NFRTD that consists of two WSMs NPs mediated by WSMs planar substrate. Due to the strong coupling of the localized plasmon modes in the NPs and nonreciprocal SPPs in the substrate, the maximum rectification ratio of NFRTD can be up to 2673, which exceeds the maximum value reported in previous works over 10 times. The contribution of the vacuum, scattering, and cross terms under the influence of WSMs NPs are investigated to explore the underlying mechanism of the high rectification ratio. Moreover, we calculate the distribution of the Green function and reflection coefficient to investigate nonreciprocal energy transfer in NFRTD. Finally, we investigate the effects of temperature and momentum-separation on the rectification performance of the NFRTD. We believe that this work can be helpful for the design of high-performance NFRTD and facilitate the application of WSMs in thermal rectification.

**Acknowledgements**

This work is supported by the National Natural Science Foundation of China (52106099, 51976173), the Shandong Provincial Natural Science Foundation (ZR2022YQ57), the Taishan Scholars Program, the Jiangsu Provincial Natural Science Foundation (BK20201204), the Basic Research Program of Taicang (TC2019JC01), and the Fundamental Research Funds for the Central Universities (D5000210779).

**References**

[1] N. Li, J. Ren, L. Wang, G. Zhang, P. Hänggi, and B. Li, Colloquium: phononics: manipulating heat flow with electronic analogs and beyond, Rev. Mod. Phys. 84, 1045 (2012).

[2] S. Basu, and M. Francoeur, Near-field radiative transfer based thermal rectification using doped silicon, Appl. Phys. Lett. 98, 113106 (2011).

[3] A. Fiorino, D. Thompson, L. Zhu, R. Mittapally, S. A. Biehs, O. Bezencenet, N. El-Bondry, S. Bansropun, P. Ben-Abdallah, and E. Meyhofer, A thermal diode based on nanoscale thermal radiation, ACS Nano 12, 5774 (2018).

[4] A. Ghanekar, J. Ji, and Y. Zheng, High-rectification near-field thermal diode using phase change periodic nanostructure, Appl. Phys. Lett. 109, 123106 (2016).

[5] J. Huang, Q. Li, Z. H. Zheng, and Y. M. Xuan, Thermal rectification based on thermochromic materials, Int. J. Heat Mass Transfer 67, 575 (2013).

[6] A. Pugsley, A. Zacharopoulos, J. Mondol, and M. Smyth, Theoretical and experimental analysis of a horizontal planar Liquid-Vapour Thermal Diode (PLVTD), Int. J. Heat Mass Transfer 144, 118660 (2019).

[7] J. D. Shen, X. L. Liu, H. He, W. T. Wu, and B. A. Liu, High-performance noncontact thermal diode via asymmetric nanostructures, J. Quant. Spectrosc. Radiat. Transfer 211, 1 (2018).

[8] L. P. Wang, and Z. M. Zhang, Thermal rectification enabled by near-field radiative heat transfer between intrinsic silicon and a dissimilar material, Nanoscale Microscale Thermophys. Eng. 17, 337 (2013).

[9] Y. Yang, S. Basu, and L. P. Wang, Vacuum thermal switch made of phase transition materials considering thin film and substrate effects, J. Quant. Spectrosc. Radiat. Transfer 158, 69 (2015).

[10] C. R. Otey, W. T. Lau, and S. H. Fan, Thermal rectification through vacuum, Phys. Rev. Lett. 104, 154301 (2010).

[11] L. X. Zhu, C. R. Otey, and S. H. Fan, Ultrahigh-contrast and large-bandwidth thermal rectification in near-field electromagnetic thermal transfer between nanoparticles, Phys. Rev. B 88, 184301 (2013).

[12] Q. Z. Li, H. Y. He, Q. Chen, and B. Song, Thin-film radiative thermal diode with large rectification, Phys. Rev. Appl. 16 014069 (2021).

[13] Q. Z. Li, H. Y. He, Q. Chen, and B. Song, Radiative thermal diode via hyperbolic metamaterials, Phys. Rev. Appl. 16, 064022 (2021).

[14] T. Villeneuve, M. Boudreau, and G. Dumas, The thermal diode and insulating potentials of a vertical stack of parallelogrammic air-filled enclosures, Int. J. Heat Mass Transfer 108, 2060 (2017).

[15] O. Ilic, N. Thomas, T. Christensen, M. Sherrott, M. Soljacic, A. Minnich, O. Miller,

and H. Atwater, Active radiative thermal switching with graphene plasmon resonators, ACS Nano, 12, 2474 (2018).

[16] B. W. Li, L. Wang, and G. Casati, Thermal diode: rectification of heat flux, Phys. Rev. Lett. 93, 184301 (2004).

[17] N. Roberts, and D. Walker, A review of thermal rectification observations and models in solid materials, Int. J. Therm. Sci. 50, 648 (2011).

[18] L. Wang, and B. W. Li, Thermal logic gates: computation with phonons, Phys. Rev. Lett. 99, 177208 (2007).

[19] T. Ruokola, T. Ojanen, and A. Jauho, Thermal rectification in nonlinear quantum circuits, Phys. Rev. B 79, 144306 (2009).

[20] Y. Wang, A. Vallabhaneni, J. N. Hu, B. Qiu, Y. P. Chen, and X. L. Ruan, Phonon lateral confinement enables thermal rectification in asymmetric single-material nanostructures, Nano Lett. 14, 592 (2014).

[21] H. D. Wang, S. Q. Hu, K. Takahashi, X. Zhang, H. Takamatsu, and J. Chen, Experimental study of thermal rectification in suspended monolayer graphene, Nat. Commun. 8, 1 (2017).

[22] M. Kasprzak, M. Sledzinska, K. Zaleski, I. Iatsunskyi, F. Alzina, S. Volz, C. Torres, and B. Graczykowski, High-temperature silicon thermal diode and switch, Nano Energy 78, 105261 (2020).

[23] R. Scheibner, M. König, D. Reuter, A. Wieck, C. Gould, H. Buhmann, and L. Molenkamp, Quantum dot as thermal rectifier, New J. Phys. 10, 083016 (2008).

[24] M. Martínez-Pérez, A. Fornieri, and F. Giazotto, Rectification of electronic heat

current by a hybrid thermal diode, Nat. Nanotechnol. 10, 303 (2015).

[25] E. Nefzaoui, K. Joulain, J. Drevillon, and Y. Ezzahri, Radiative thermal rectification using superconducting materials, Appl. Phys. Lett. 104, 103905 (2014).

[26] I. Latella, P. Ben-Abdallah, and M. Nikbakht, Radiative thermal rectification in many-body systems, Phys. Rev. B 104, 045410 (2021).

[27] X. H. Wu, C. J. Fu, and Z. M. Zhang, Near-field radiative heat transfer between two $\alpha$-MoO$_3$ biaxial crystals, J. Heat Transfer 142, 072802 (2020).

[28] X. H. Wu, and C. J. Fu, Near-field radiative heat transfer between uniaxial hyperbolic media: role of volume and surface phonon polaritons, Journal of J. Quant. Spectrosc. Radiat. Transfer 258, 107337 (2021).

[29] X. H. Wu, and C. J. Fu, Near-field radiative modulator based on dissimilar hyperbolic materials with in-plane anisotropy, Int. J. Heat Mass Transfer 168, 120908 (2021).

[30] R. Y. Liu, C. L. Zhou, Y. Zhang, Z. Cui, X. H. Wu, and H. L. Yi, Near-field radiative heat transfer in hyperbolic materials, Int. J. Extreme Manuf. 4, 032002 (2022).

[31] Y. Hu, Y. S. Sun, Z. H. Zheng, J. L. Song, K. Z. Shi, and X. H. Wu, Rotation-induced significant modulation of near-field radiative heat transfer between hyperbolic nanoparticles, Int. J. Heat Mass Transfer 189, 122666 (2022).

[32] J. Dong, W. J. Zhang, and L. H. Liu, Nonreciprocal thermal radiation of nanoparticles via spin-directional coupling with reciprocal surface modes, Appl. Phys. Lett. 119, 021104 (2021).

[33] M. G. Luo, J. Dong, J. M. Zhao, L. H. Liu, and M. Antezza, Radiative heat transfer between metallic nanoparticle clusters in both near field and far field, Phys. Rev. B 99, 134207 (2019).

[34] D. Xu, J. Zhao, and L. Liu, Near-field thermal radiation of gradient refractive index slab: Internal polaritons, Appl. Phys. Lett. 119, 141106 (2021).

[35] A. Ott, S. A. Biehs, and P. Ben-Abdallah, Anomalous photon thermal Hall effect, Phys. Rev. B 101, 241411 (2020).

[36] S. A. Biehs, R. S. Messina, P. Venkataram, A. W. Rodriguez, J. C. Cuevas, and P. Ben-Abdallah, Near-field radiative heat transfer in many-body systems, Rev. Mod. Phys. 93, 025009 (2021).

[37] K. Z. Shi, Z. Y. Chen, X. N. Xu, J. Evans, and S. L. He, Optimized colossal near-field thermal radiation enabled by manipulating coupled plasmon polariton geometry, Adv. Mater. 33, 2106097 (2021).

[38] K. Z. Shi, R. Liao, G. J. Cao, F. L. Bao, and S. L. He, Enhancing thermal radiation by graphene-assisted hBN/SiO$_2$ hybrid structures at the nanoscale, Opt. Express 26, A5911 (2018).

[39] K. Z. Shi, Z. Y. Chen, Y. X. Xing, J. X. Yang, X. N. Xu, S. Evans, S. L. He, Near-field radiative heat transfer modulation with an ultrahigh dynamic range through mode mismatching, Nano Lett. 22, 7753 (2022).

[40] K. Z. Shi, Y. C. Sun, Z. Y. Chen, N. He, F. L. Bao, J. Evans, and S. L. He, Colossal enhancement of near-field thermal radiation across hundreds of nanometers between millimeter-scale plates through surface plasmon and phonon polaritons

coupling, Nano Lett. 19, 8082 (2019).

[41] K. Z. Shi, F. L. Bao, and S. L. He, Enhanced near-field thermal radiation based on multilayer graphene-hBN heterostructures, ACS Photonics 4, 971 (2017).

[42] K. Z. Shi, F. L. Bao, N. He, and S. L. He, Near-field heat transfer between graphene-Si grating heterostructures with multiple magnetic-polaritons coupling, Int. J. Heat Mass Transfer 134, 1119 (2019).

[43] M. J. He, H. Qi, Y. T. Ren, Y. J. Zhao, and M. Antezza, Magnetoplasmonic manipulation of nanoscale thermal radiation using twisted graphene gratings, Int. J. Heat Mass Transfer 150, 119305 (2020).

[44] M. J. He, H. Qi, Y. T. Ren, Y. J. Zhao, and M. Antezza, Graphene-based thermal repeater, Appl. Phys. Lett. 115, 263101 (2019).

[45] Y. Zhang, H. L. Yi, H. P. Tan, and M. Antezza, Giant resonant radiative heat transfer between nanoparticles, Phys. Rev. B 100, 134305 (2019).

[46] Y. Zhang, M. Antezza, H. L. Yi, and H. P. Tan, Metasurface-mediated anisotropic radiative heat transfer between nanoparticles, Phys. Rev. B 100, 085426 (2019).

[47] Z. H. Zheng, X. L. Liu, A. Wang, and Y. M. Xuan, Graphene-assisted near-field radiative thermal rectifier based on phase transition of vanadium dioxide ($VO_2$), Int. J. Heat Mass Transfer 109, 63 (2017).

[48] E. Moncada-Villa, and J. Cuevas, Normal-metal–superconductor near-field thermal diodes and transistors, Phys. Rev. Appl. 15, 024036 (2021).

[49] Y. Liu, Y. P. Tian, F. Q. Chen, A. Caratenuto, X. J. Liu, M. Antezza, and Y. Zheng, Ultrahigh-rectification near-field radiative thermal diode using infrared-

transparent film backsided phase-transition metasurface, Appl. Phys. Lett. 119, 123101 (2021).

[50] P. Doyeux, S. Gangaraj, G. Hanson, and M. Antezza, Giant interatomic energy-transport amplification with nonreciprocal photonic topological insulators, Phys. Rev. Lett. 119, 173901 (2017).

[51] A. Ott, R. Messina, P. Ben-Abdallah, and S. A. Biehs, Radiative thermal diode driven by nonreciprocal surface waves, Appl. Phys. Lett. 114, 163105 (2019).

[52] B. Zhao, C. Guo, C. A. C. Garcia, P. Narang, and S. H. Fan, Axion-field-enabled nonreciprocal thermal radiation in Weyl semimetals, Nano Lett. 20, 1923 (2020).

[53] C. Guo, B. Zhao, D. Huang, and S. H. Fan, Radiative thermal router based on tunable magnetic Weyl semimetals, ACS Photonics 7, 3257 (2020).

[54] J. Z. Wu, H. T. Li, C. J. Fu, and X. H. Wu, High quality factor nonreciprocal thermal radiation in a Weyl semimetal film via the strong coupling between tamm plasmon and defect mode, Int. J. Therm. Sci. 184, 107902 (2023).

[55] Y. S. Sun, Y. Hu, K. Z. Shi, J. H. Zhang, D. D. Feng, and X. H. Wu, Negative differential thermal conductance between Weyl semimetals nanoparticles through vacuum, Phys. Scr. 97, 095506 (2022).

[56] Z. M. Zhang, and L. X. Zhu, Nonreciprocal thermal photonics for energy conversion and radiative heat transfer, Phys. Rev. Appl. 18, 027001 (2022).

[57] J. Wu, B. Y. Wu, Z. M. Wang, and X. H. Wu, Strong nonreciprocal thermal radiation in Weyl semimetal-dielectric multilayer structure, Int. J. Therm. Sci. 181, 107788 (2022).

[58] B. Zhao, J. H. Wang, Z. X. Zhao, C. Guo, Z. F. Yu, and S. H. Fan, Nonreciprocal thermal emitters using metasurfaces with multiple diffraction channels, Phys. Rev. Appl. 16, 064001 (2021).

[59] X. H. Wu, H. Y. Yu, F. Wu, and B. Y. Wu, Enhanced nonreciprocal radiation in Weyl semimetals by attenuated total reflection, AIP Adv. 11, 075106 (2021).

[60] G. M. Tang, J. Chen, and L. Zhang, Twist-induced control of near-field heat radiation between magnetic Weyl semimetals, ACS Photonics 8, 443 (2021).

[61] N. Armitage, E. Mele, and A. Vishwanath, Weyl and Dirac semimetals in three-dimensional solids, Rev. Mod. Phys. 90, 015001 (2018).

[62] Z. Yu, X. Li, T. Lee, and H. Iizuka, Near-field radiative heat transfer between Weyl semimetal multilayers, Int. J. Heat Mass Transfer 197, 123339 (2022).

[63] G. D. Xu, J. Sun, and H. M. Mao, Near-field radiative thermal modulation between Weyl semimetal slabs, J. Quant. Spectrosc. Radiat. Transfer 253, 107173 (2020).

[64] Z. Q. Yu, X. P. Li, T. Lee, and H. Iizuka, Near-field radiative heat transfer in three-body Weyl semimetals, Opt. Express 30, 31584 (2022).

[65] X. H. Wu, C. J. Fu, and Z. M. Zhang, Influence of hBN orientation on the near-field radiative heat transfer between graphene/hBN heterostructures, J. Photonics Energy 9, 032702 (2018).

[66] R. Messina, M. Tschikin, S. A. Biehs, and P. Ben-Abdallah, Fluctuation-electrodynamic theory and dynamics of heat transfer in systems of multiple dipoles, Phys. Rev. B 88, 104307 (2013).

[67] P. Ben-Abdallah, S. A. Biehs, and K. Joulain, Many-body radiative heat transfer

theory, Phys. Rev. Lett. 107, 114301 (2011).

[68] J. Soh, F. De Juan, M. Vergniory, N. Schröter, M. Rahn, D. Yan, J. Jiang, M. Bristow, P. Reiss, and J. Blandy, Ideal Weyl semimetal induced by magnetic exchange, Phys. Rev. B 100, 201102 (2019).

[69] V. Asadchy, C. Guo, B. Zhao, and S. H. Fan, Sub‐wavelength passive optical isolators using photonic structures based on Weyl semimetals, Adv. Opt. Mater. 8, 2000100 (2020).

[70] K. Wang, and L. Gao, High-efficient photonic thermal rectification with magnetocontrollability, ES Energy Environ. 7, 12 (2019).

[71] A. Ott, and S. A. Biehs, Thermal rectification and spin-spin coupling of nonreciprocal localized and surface modes, Phys. Rev. B 101, 155428 (2020).

[72] A. Manjavacas, and F. Abajo, Radiative heat transfer between neighboring particles, Phys. Rev. B 86, 075466 (2012).

[73] R. Ekeroth, A. García-Martín, and J. Cuevas, Thermal discrete dipole approximation for the description of thermal emission and radiative heat transfer of magneto-optical systems, Phys. Rev. B 95, 235428 (2017).

[74] S. Albaladejo, R. Gómez-Medina, L. Froufe-Pérez, H. Marinchio, R. Carminati, J. Torrado, G. Armelles, A. García-Martín, and J. Sáenz, Radiative corrections to the polarizability tensor of an electrically small anisotropic dielectric particle, Opt. Express 18, 3556-3567 (2010).

[75] J. Dong, J. M. Zhao, and L. H. Liu, Long-distance near-field energy transport via propagating surface waves, Phys. Rev. B 97, 075422 (2018).

[76] Y. Zhang, C. L. Zhou, H. L. Yi, and H. P. Tan, Radiative thermal diode mediated by nonreciprocal graphene plasmon waveguides, Phys. Rev. Appl. 13, 034021 (2020).

[77] Y. Yang, S. Basu, and L. P. Wang, Radiation-based near-field thermal rectification with phase transition materials, Appl. Phys. Lett. 103, 163101 (2013).

[78] M. Q. Yuan, Y. Zhang, S. H. Yang, C. L. Zhou, and H. L. Yi, Near-field thermal rectification driven by nonreciprocal hyperbolic surface plasmons, Int. J. Heat Mass Transfer 185, 122437 (2022).

[79] R. J. William, Transient shutdown analysis of low-temperature thermal diodes, NASA Technical Paper 1369, N79 (1979).

[80] R. Williams, Investigation of a cryogenic thermal diode, AIAA, 78 (1978).

[81] G. T. Colwell, Prediction of cryogenic heat pipes performances, Final Report, N77 (1975).

[82] A. Ott, P. Ben-Abdallah, and S. A. Biehs, Circular heat and momentum flux radiated by magneto-optical nanoparticles, Phys. Rev. B 97, 205414 (2018).

[83] I. Latella, and P. Ben-Abdallah, Giant thermal magnetoresistance in plasmonic structures, Phys. Rev. Lett. 118, 173902 (2017).

[84] R. Ekeroth, P. Ben-Abdallah, J. Cuevas, and A. García-Martín, Anisotropic thermal magnetoresistance for an active control of radiative heat transfer, ACS Photonics 5, 705 (2018).

[85] P. Ben-Abdallah, Photon thermal Hall effect, Phys. Rev. Lett. 116, 084301 (2016).

[86] R. Messina, S. A. Biehs, and P. Ben-Abdallah, Surface-mode-assisted amplification

of radiative heat transfer between nanoparticles, Phys. Rev. B 97, 165437 (2018).

[87] K. Joulain, J. P. Mulet, F. Marquier, R. Carminati, and J. J. Greffet, Surface electromagnetic waves thermally excited: Radiative heat transfer, coherence properties and Casimir forces revisited in the near field, Surf. Sci. Rep. 57, 59 (2005).

[88] P. Ben-Abdallah, Multitip near-field scanning thermal microscopy, Phys. Rev. Lett. 123, 264301 (2019).

[89] D. Thompson, L. X. Zhu, E. Meyhofer, and P. Reddy, Nanoscale radiative thermal switching via multi-body effects, Nat. Nanotechnol. 15, 99 (2020).